\documentclass[12pt]{article}
\usepackage{amsthm}
\newtheorem{definition}{Definition}
\usepackage{float}
\usepackage{array}
\usepackage{booktabs} % For better table lines
\usepackage{ragged2e} % For text alignment
\usepackage{graphicx}
\usepackage{amsfonts}
\usepackage{amsmath}
\usepackage[utf8]{inputenc}
\usepackage[english]{babel}
\usepackage{amssymb,amsmath,amsthm}
\usepackage{url}
\usepackage{lineno}
\usepackage{fontspec}
\usepackage[colorlinks=true, 
	breaklinks=true,
	citecolor=blue,
    colorlinks=true
]{hyperref}
\hypersetup{
    allcolors=blue, % Makes all links (including citations, refs) blue
    linkcolor=blue,     % Internal links (sections, pages)
    citecolor=blue,     % Citations
    urlcolor=blue,      % URLs
    filecolor=blue,     % File links
    anchorcolor=blue,   % Anchor text
}
\usepackage{setspace}
\usepackage{geometry}
\usepackage{xcolor}

\usepackage[super]{natbib}
\newtheorem{theorem}{Theorem}

\usepackage{scicite}

\usepackage{times}

\usepackage{newtxtext}
\usepackage{newtxmath}

\title{Finding the Best Route  During the Pandemic Disease} 

\author
{Amirsadegh Mirgalooyebayat,$^{1\ast}$ Farzad Didehvar$^{1\ast}$\\
\\
\normalsize{$^{1}$Department of Mathematics and Computer Science, Amirkabir University of Technology, Tehran, Iran}
\\
\normalsize{$^\ast$E-mail: amir.bayat@aut.ac.ir,  didehvar@aut.ac.ir}
}

\date{August 2025}

\begin{document} 

\baselineskip24pt

\maketitle

\section*{Contracts of Interest}
The authors declare they have no conflicts of interest related to this work to disclose.

\section*{Abstract}

  This article presents a mathematical model for identifying the safest travel routes during a pandemic by minimizing disease contraction risks, such as COVID-19. We formulate this as the \textsc{Least Infection Probability Path} (LIPP) problem, which optimizes routes between two nodes in a transportation network based on minimal disease transmission probability. Our model evaluates risk factors including environmental density, likelihood of encountering carriers, and exposure duration across multiple transportation modes (walking, subway, BRT, buses, and cars).
  
  The probabilistic framework incorporates additional variables such as ventilation quality, activity levels, and interpersonal distances to estimate transmission risks. Applied to Tehran's transportation network using routing applications (\textit{Neshan} and \textit{Balad}), our model demonstrates that combined pedestrian-subway-BRT routes exhibit significantly lower infection risks compared to car or bus routes, as illustrated in our case study of peak-hour travel between Sadeghiyeh Square and Amirkabir University.
  
  We develop a practical routing algorithm suitable for integration with existing navigation software to provide pandemic-aware path recommendations. Potential future extensions include incorporating additional variables like waiting times and line changes, as well as adapting the model for other infectious diseases. This research offers a valuable tool for urban travelers seeking to minimize infection risks during pandemic conditions.

Key Words:

Covid-19, pandemic, mathematical model, routing applications, routing, Least Infection Probability Path Problem.

\section*{Introduction}
In a 2015 TED Talk, Bill Gates warned that a highly infectious virus, not war, would likely cause the next major global catastrophe, criticizing the lack of preparedness for such an outbreak. Recently, he reiterated his concerns, estimating a 10–15\% chance of another pandemic within four years due to ongoing political divisions and weak health systems. Despite COVID-19’s lessons, he emphasizes the need for better disease surveillance, faster vaccine development, and stronger international cooperation. Even if these measures resolve the pandemic threat, the risk of epidemics persists. These facts underscore the critical importance of epidemic preparedness in global health and security.\citep{BillGatesRecentlyPrediction,BillGatesTedTalk}

The Covid-19 pandemic (2019–2021) revived interest in understanding and managing infectious disease outbreaks. Mathematical modeling has been a key tool for predicting and analyzing disease spread. In 2013, Huppert and Katriel reviewed the application of mathematical models in infectious disease epidemics, emphasizing their role in estimating parameters like the initial reproduction number ($R_0$) and evaluating interventions such as vaccination and social distancing.\citep{huppert2013mathematical}

During the pandemic, mathematical modeling became essential for estimating transmission dynamics. Lelieveld et al. (2020) investigated Covid-19 transmission through aerosol particles in indoor environments, using models to calculate viral particle concentrations and infection risks based on room size, occupancy, and ventilation rates.\citep{lelieveld2020model} Guzman (2021) further explored the persistence of virus-laden particles in the air, highlighting their ability to travel up to 2 meters and remain airborne for hours.\citep{guzman2021overview}

The role of ventilation in controlling virus spread was emphasized by Li et al. (2021), who analyzed an outbreak in a poorly ventilated restaurant, demonstrating the potential for airborne transmission in enclosed spaces.\citep{li2021probable} Vuorinen et al. (2020) used numerical simulations to model aerosol transmission, assessing risks based on ventilation, occupancy, and activity levels.\citep{vuorinen2020modelling} Anderson et al. (2020) also studied aerosol transmission, focusing on how airflow, humidity, and temperature affect viral particle distribution indoors.\citep{anderson2020consideration}

Environmental factors like wind and humidity were examined by Feng et al. (2020), who used simulations to assess their impact on social distancing effectiveness.\citep{feng2020influence} Hassan et al. (2021) developed a mathematical model to predict Covid-19 spread in Texas, incorporating factors such as population density and mobility patterns.\citep{hassan2023mathematical} Miralles-Pechuán et al. (2020) combined the SEIR model with Deep Q-learning and genetic algorithms to optimize quarantine and travel restriction policies.\citep{miralles2020methodology}

Transmission risks in confined spaces, such as passenger cars and public transportation, were also studied. Sarhan et al. (2022) modeled virus spread in cars, considering ventilation, mask usage, and passenger positioning.\citep{sarhan2022numerical} Zhao et al. (2024) investigated droplet dispersion in city buses, showing that droplets could spread throughout the compartment within seconds.\citep{zhao2024characteristics} Gkiotsalitis and Cats (2020) reviewed public transportation adaptations during the pandemic, highlighting challenges in balancing safety and efficiency.\citep{gkiotsalitis2021public} Marra et al. (2022) explored changes in route selection, noting a preference for less crowded routes to minimize exposure.\citep{marra2022impact}

Recent advancements in 2024–2025 have further refined our understanding of disease transmission in urban environments. Blackmore and Lloyd-Smith (2024) developed mathematical models to quantify disease spread via mobility networks, emphasizing the role of transportation hubs in accelerating outbreaks.\citep{blackmore2024transoceanic} Dynamic route planning has also evolved, with Local Eyes (2024) demonstrating how real-time adjustments to transportation routes can reduce exposure risks by 20\% through AI-driven optimization of traffic and congestion data.\citep{Localeyes} Public-private collaborations like the BRIDGE Alliance (2025) now leverage AI and cross-sector data sharing to enhance global disease surveillance, addressing gaps in pandemic preparedness.\citep{HealthcareSystems} Additionally, improved SEIR models, such as the SEIA framework (2024), now account for asymptomatic transmission in subways—a critical factor in urban epidemic control.\citep{zhou2024prevention}

This article addresses the problem of identifying urban routes with the lowest probability of disease transmission. Previous studies have explored related challenges, such as optimizing vehicle routing for contactless delivery\citep{chen2020vehicle} and medical waste collection.\citep{eren2021safe} Pacheco and Laguna (2020) developed methods for optimizing face shield delivery during the pandemic,\citep{pacheco2020vehicle} while Păcurar et al. (2021) applied the traveling salesman problem to find safe tourist routes.\citep{puacurar2021tourist}

The proposed approach uses mathematical modeling to calculate infection probabilities for different transportation modes, including walking, subway, BRT, bus, and car. The model is applied to real-world scenarios using routing applications like \textit{Neshan} and \textit{Balad} to evaluate routes between Sadeghieh Square and Amirkabir University in Tehran. By analyzing these routes, the study identifies the safest options and provides insights for future research and policy development.

\section*{Methods}

Our methodological framework integrates mathematical modeling of COVID-19 transmission with transportation network analysis to assess infection risks across urban routes. The approach combines theoretical foundations with empirical parameter estimation and practical algorithm implementation.

\subsection*{Conceptual Framework}

The model architecture decomposes complex urban routes into five distinct microenvironment types, each with unique transmission characteristics. These microenvironments include pedestrian pathways, subway carriages, BRT vehicles, city buses, and private cars. Each microenvironment is characterized by its geometric properties (dimensions and layout) and ventilation parameters (airflow rates and open/closed space configurations). For example, subway carriages have different ventilation characteristics than pedestrian walkways, leading to distinct transmission dynamics.

The exposure assessment component quantifies both the duration of potential exposures and the exposure intensity based on activity levels. The transmission dynamics module incorporates disease-specific aerosol propagation models that account for viral load and transmission probability, using COVID-19 specific parameters from empirical studies. The dimensions of the environment, the duration of exposure (time spent in the environment), and the activity level of individuals—which determines the air inhalation rate in liters per hour—are used as input values to calculate the probability of infection.\citep{covid-19} Additional parameters, such as ventilation rate or window usage, also influence these calculations; however, we omit their variability by introducing simplifying assumptions. Using this probability for different environments and the specified mathematical model, we then estimate the coefficients of the model for Covid-19.

The route optimization component develops algorithms for minimal-risk path selection that balance infection risk against travel time and convenience. This allows comparison of different route options (e.g., subway vs. bus routes) for the same origin-destination pair.

\subsection*{Probability Model Development}

\subsubsection*{Composite Route Risk}
\begin{figure}[h!]
    \centering
    \includegraphics[width=\linewidth]{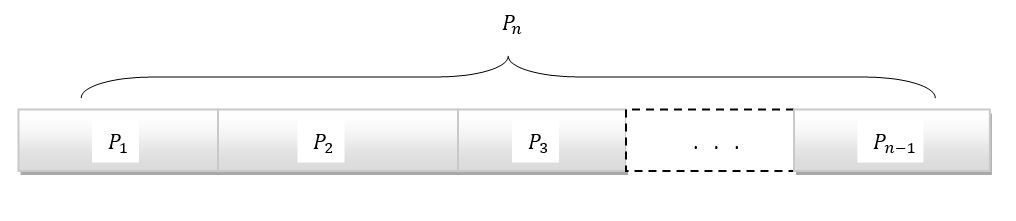}
    \caption{Diagram of a combined route showing different microenvironment segments}
    \label{fig:fig1_1}
\end{figure}

In our study, each path can be divided into several segments, as shown in Figure~\ref{fig:fig1_1}. The probability $P_n$ of contracting COVID-19 along route $n$ (with $n-1$ segments) follows a multiplicative risk model as shown in Equation~\ref{eq:composite_risk}.

\begin{theorem}
The probability $P_n$ of contracting COVID-19 along route $n$ consisting of $n-1$ independent segments is:
\begin{equation}
\label{eq:composite_risk}
P_n = 1 - \prod_{i=1}^{n-1} (1 - P_i)
\end{equation}
where $P_i$ is the transmission probability for segment $i$.
\end{theorem}

This formulation accounts for the independent probabilities of infection across consecutive route segments. For instance, a route combining subway and walking segments would integrate the transmission risks from both environments.

\subsubsection*{Segment Risk Calculation}

To evaluate the infection risk in each segment, we consider three parameters: environmental density (\(\rho\)), the probability that an individual is a carrier (\(p\)), and the duration of exposure (\(t\)). Here, \textit{environmental density} refers to the number of people per unit area in the studied environment.

The probability of contracting the disease for specific environmental density and carrier probability is a function from the time interval ($\text{Int}(\mathbb{R}^+)$) to $[0,1]$.

\begin{definition}[Infection Probability Function]
\label{def:infection_prob}
The infection probability function for given environmental density $\rho$ and carrier probability $p$ maps time intervals to probability values:
\[
\hat{f}_{\rho,p}: \text{Int}(\mathbb{R}^+) \rightarrow [0,1]
\]
where $\text{Int}(\mathbb{R}^+)$ denotes the set of all time intervals $[t_1,t_2]$ for $t_1, t_2 \in \mathbb{R}^+$.
\end{definition}

Similar to the composite path risk scenario, an individual remaining in the environment during the interval $[0,a+b]$ experiences equivalent exposure to consecutive intervals $[0,a]$ and $[a,a+b]$. Because the probability depends only on interval duration (due to the time-homogeneous nature of the process), it follows that $\hat{f}_{\rho,p}([0,b]) = \hat{f}_{\rho,p}([a,a+b])$. Theorem~\ref{th:time_function} provides the probability of contracting the disease as a function of the time interval.
\begin{theorem}
If an individual is exposed to the environment during the time interval $[0,a+b]$, the infection probability as a function of time satisfies:
\begin{equation}
\label{eq:time_function0}
    \hat{f}_{\rho,p}([0,a+b]) = \hat{f}_{\rho,p}([0,a]) + \hat{f}_{\rho,p}([0,b]) - \hat{f}_{\rho,p}([0,a])\hat{f}_{\rho,p}([0,b])
\end{equation}
\label{th:time_function}
\end{theorem}

Since the infection probability depends only on exposure duration (due to the time-homogeneity of the process), the interval-based probability function $\hat{f}_{\rho,p}([0,t])$ is equivalent to the duration-based function $f_{\rho,p}(t)$. This equivalence is expressed as:
\[
\hat{f}_{\rho,p}([0,t]) = f_{\rho,p}(t),
\]
where
\[
f_{\rho,p}: \mathbb{R}^+ \to [0,1]
\]
is a function mapping from the set of positive real numbers (representing exposure durations) to probabilities in the unit interval $[0,1]$.

In Theorem~\ref{th:time_natural}, we prove that the probability of disease transmission over time follows a natural-exponential relationship under constant exposure conditions.
\begin{theorem}
If the duration of time an individual is in the environment is equal to $t$ and $k$ is a natural and constant number, for all arbitrary and natural numbers $n$, the probability of getting a disease in terms of time can be calculated from the following equation:
\begin{equation}
	f_{\rho,p}(kf^{-1} (\frac{1}{n}))=1-(\frac{n-1}{n})^k ;k,n \in \mathbb {N}
\end{equation}
\label{th:time_natural}
\end{theorem}

In Theorem \ref{th:time_function3_0}, we establish that the infection probability follows a compound risk model under partitioned exposure intervals.
\begin{theorem}
If the duration of time an individual is in the environment is equal to $t$ and $k$ is a natural and constant number, the probability of getting a disease in terms of time can be calculated from the following equation:
\begin{equation}
	f_{\rho,p}(t)=1-(1-f_{\rho,p}(\frac{t}{k}))^k
	\label{eq:time_function3_0}
\end{equation}
\label{th:time_function3_0}
\end{theorem}

The probability of infection as a function of time is calculated in Equation \ref{eq:time_function4_0}.
\begin{theorem}
The probability of infection as a function of time is given by the following relationship, where c is a time-dependent constant:
\begin{equation}
	f(t)=1-e^{ct}
	\label{eq:time_function4_0}
\end{equation}
\end{theorem}

Since the infection probability depends on three independent parameters---environmental density ($\rho$), carrier probability ($p$), and exposure duration ($t$)---we can express the function from Eq.~\eqref{eq:time_function4_0} as:
\begin{equation}
	F(\rho, p, t) = 1 - e^{-\kappa(\rho, p) t}
	\label{eq:eq0}
\end{equation}
where \(\kappa\) is a function dependent on \(\rho\) and \(p\).

\begin{figure}[h]
    \centering
    \includegraphics[width=0.8\linewidth]{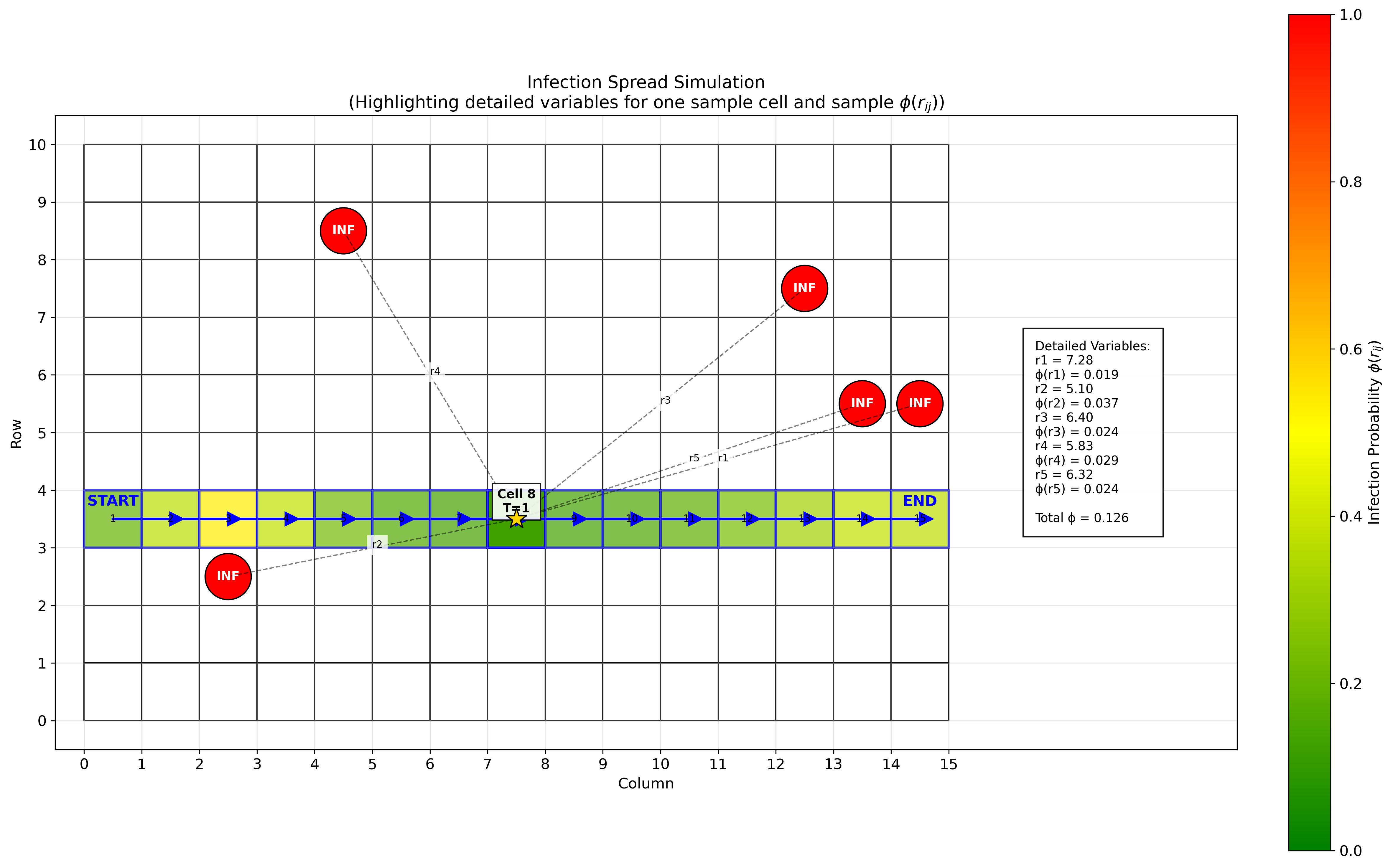}
    \caption{Infection risk ($\phi(r_{ij})$) for an uninfected individual (blue path) exposed to *n* infected individuals (red). The gold star highlights a sample cell’s risk variables ($r_(ij), T_j, \phi$).}
    \label{fig:Infection_Spread_Simulation}
\end{figure}

To compute \(\kappa(\rho, p)\) we model a 2D environment of size $m \times l$ with $n$ randomly placed infected individuals. According to Figure~\ref{fig:Infection_Spread_Simulation} an uninfected individual traverses this environment, and we compute their probability of infection. The model defines $T_j$ as the duration spent in cell $j$ and $r_{ij}$ as the distance to an infected individual $i$. 

We model the infection probability using a distance-dependent transmission function ($\phi(r_{ij})$) with linear dependence on exposure time ($\propto T_j$). Based on these assumptions, we derive the following equations:

\begin{enumerate}
    \item Cell infection probability:
    \begin{equation}
        g\left(\sum_{i=1}^n \phi(r_{ij})\right)T_j
        \label{eq:cell_prob}
    \end{equation}
    
    \item Path survival probability (independent cells):
    \begin{equation}
        \prod_{j=1}^m \left[1 - g\left(\sum_{i=1}^n \phi(r_{ij})\right)T_j\right]
    \end{equation}
    
    \item Total infection probability:
    \begin{equation}
        h(k) = 1 - \prod_{j=1}^m \left[1 - g\left(\sum_{i=1}^n \phi(r_{ij})\right)T_j\right]
        \label{eq:total_prob}
    \end{equation}
\end{enumerate}

Since both \( F(\rho, p, t) \) and \( h(k) \) represent infection risk probabilities, we equate them:
\begin{equation}
    F(\rho, p, t) = h(k) \implies 1 - e^{-\kappa(\rho,p)t} = 1 - \prod_{j=1}^m \left(1 - g\left(\sum_{i=1}^n \phi(r_{ij})\right)T_j\right)
    \label{eq:equality}
\end{equation}

We hypothesize that the initial equation should be formulated to derive the following expression:
\begin{equation}
    \kappa(\rho,p) = \frac{1}{t} \sum_{j=1}^m P\left(\ln\left(1 - \sum_{i=1}^n \frac{\phi(r_{ij})}{m}\right)\right); \quad P(t) = 1 - e^t
    \label{eq:eq1}
\end{equation}

Substituting \( P(t) \) from Equation \eqref{eq:eq1} yields the simplified relationship:
\begin{equation}
    \kappa(\rho,p) = \frac{1}{t}\sum_{j=1}^m \sum_{i=1}^n \frac{\phi(r_{ij})t}{m} = 
    \sum_{j=1}^m \sum_{i=1}^n \frac{\phi(r_{ij})}{m}
    \label{eq:eq2}
\end{equation}

In this formulation, the coefficient \( \kappa(\rho,p) \) becomes time-independent. We assume that the distance between the susceptible individual and infected individuals remains constant throughout the exposure period. Under this assumption, Equation~\eqref{eq:eq0} simplifies to:

\begin{equation}
    F(\rho, p, t) = 1 - \exp\left(-\sum_{i=1}^n \phi(r_{ij})\right),
    \label{eq:eq6}
\end{equation}

where \( n \) represents the number of infected individuals in the environment.

In this article, we assume an inverse-square relationship between distance and infection probability; thus, Equation~\eqref{eq:eq6} simplifies to:

\begin{equation}
    F(\rho, p, t) = 1 - \exp\left(-\sum_{i=1}^n \frac{k t}{r_i^2}\right),
    \label{eq:eq3}
\end{equation}

To estimate the transmission coefficient \( k \), we consider an experimental scenario where two individuals (one infected) are placed in a 1\,m $\times$ 1\,m room for duration \( t \), maintaining a fixed separation distance. Setting \( n = 1 \) in Equation~\eqref{eq:eq3} yields:

\begin{equation}
    F(\rho, p, t) = 1 - \exp\left(-\frac{k t}{r^2}\right).
    \label{eq:eq4}
\end{equation}

Given an empirical infection probability \( F(\rho, p, t) \), we can solve for \( k \):

\begin{equation}
    k = -\frac{r^2}{t} \ln\left(1 - F(\rho, p, t)\right).
    \label{eq:eq5}
\end{equation}

\begin{figure}[h!]
    \centering
    \includegraphics[width=0.8\linewidth]{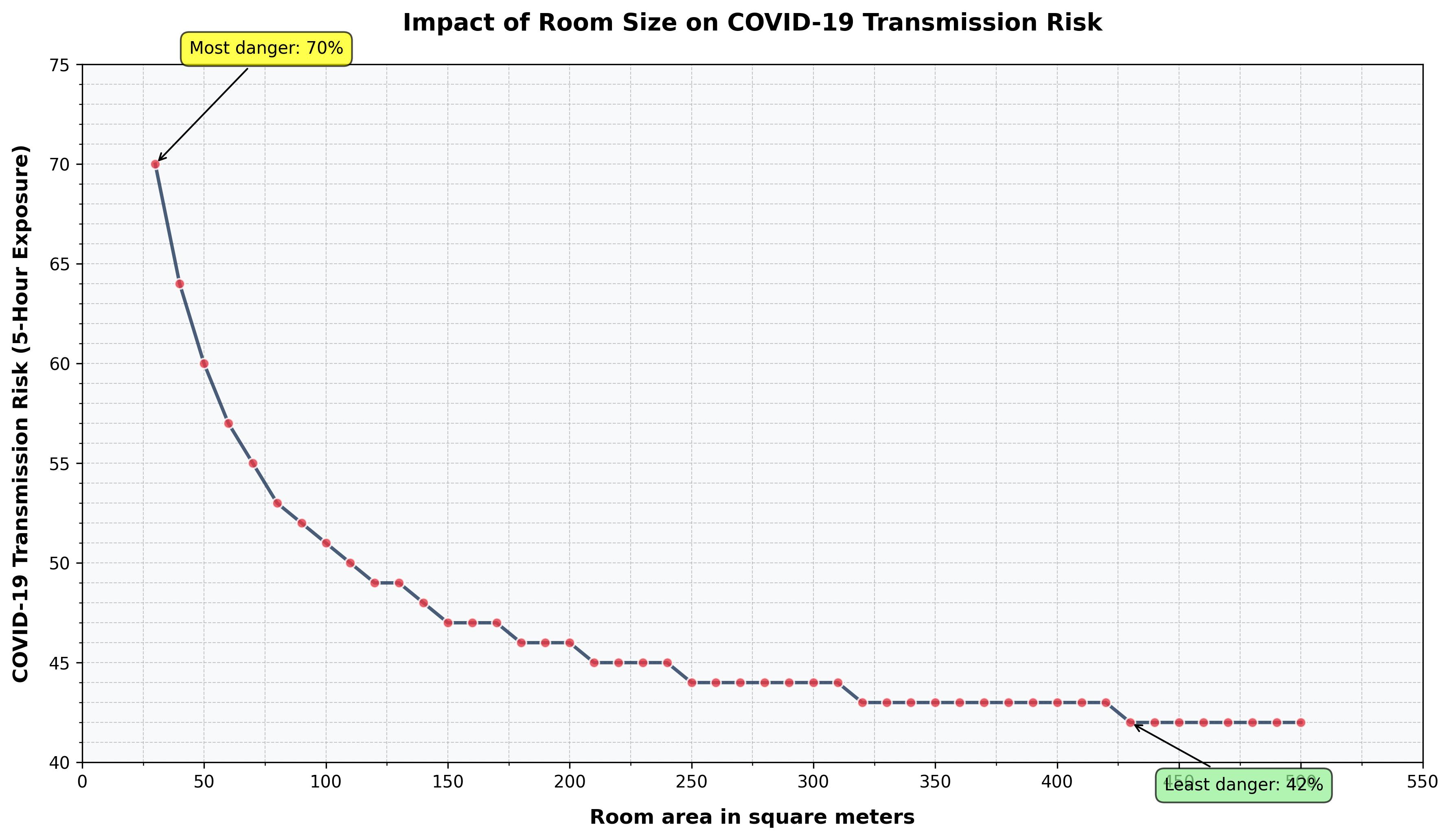}
    \caption{Probability of COVID-19 infection after 5 hours of exposure in low-activity environments as a function of room area.}
    \label{fig:fig1_0}
\end{figure}

We calibrated \( k \) using COVID-19 transmission data from \textit{COVID-19 Risk Calculator}\citep{covid-19}, analyzing enclosed spaces of varying sizes with one infected individual at an average interpersonal distance of 6 feet (1.83\,m). As shown in Figure~\ref{fig:fig1_0}, the infection probability asymptotically approaches a constant value for large room areas (e.g., 500\,m\textsuperscript{2}), suggesting that open-air transmission risks converge to those of sufficiently large enclosed spaces.

Our analysis reveals that \( k \) depends primarily on activity levels, remaining approximately constant for fixed exertion. We therefore generalize \( k \) to a function \( k(E) \), where \( E \) denotes the inhalation rate in the environment. As demonstrated in arxiv version of this article \citep{mirgalooyebayat2024finding}, \( k(E) \) exhibits a linear dependence on \( E \), implying that transmission probabilities follow an exponential survival model for each environmental segment:

\begin{equation}
P_i = 1 - \exp\left(-\lambda_i t_i\right)
\label{eq:segment_risk}
\end{equation}

where $\lambda_i$ is the hazard rate (infections per hour) and $t_i$ is the exposure duration in hours. The environment-specific hazard rate incorporates multiple transmission factors:

\begin{equation}
\lambda_i = \frac{(a_i E + b_i)n_i}{r_{i,mean}^2}
\label{eq:hazard_rate}
\end{equation}

The parameters include activity factors (air inhalation rate $E$ and environment coefficients $a_i$, $b_i$) and exposure factors (expected infected individuals $n_i$ and mean interpersonal distance $r_{i,mean}$). The environmental coefficients $a_i$, $b_i$, presented in Table \eqref{tab:env_coeffs}, correspond to various environments.\citep{mirgalooyebayat2024finding}

\begin{table}[H]
\centering
\caption{Transmission model coefficients by environment}
\label{tab:env_coeffs}
\begin{tabular}{@{}lccp{5cm}@{}}
\toprule
Environment & $a_i$ & $b_i$ \\
\midrule
Pedestrian & $-0.001439$ & $0.714554$ \\
Subway & $-0.001801$ & $0.824029$ \\
BRT & $-0.001491$ & $0.388753$ \\
City Bus & $-0.002203$ & $0.565659$ \\
Car & -- & $-2.729480$ \\
\bottomrule
\end{tabular}
\end{table}

\subsection*{Parameter Estimation}

\subsubsection*{Infected Population Estimation}

 The prevalence rate can be calculated as the division of active cases by the total resident population. Using this method, the prevalence rate was calculated as:

\[
\rho = \frac{\text{Active cases}}{\text{Residents}}
\]

This prevalence rate was then applied to vehicle capacities to estimate the expected number of infected individuals:

\begin{equation}
n_i = \rho \times \text{Capacity}_i
\label{eq:prevalence_ratio}
\end{equation}

During the peak prevalence period in September 2021, the number of infected individuals in each microenvironment was estimated using Iranian Ministry of Health reports. With an estimated 740,000 active cases among 84,500,000 residents, $\rho$ was calculated as 0.008655, which is the value used in this study \citep{IranPopulation,covidIranShahrivar1400}. By implementing this ratio in Equation \ref{eq:prevalence_ratio}, the expected number of infected individuals in each environment can be calculated.

\subsubsection*{Microenvironment Geometry}

Mean interpersonal distances were calculated using a comprehensive geometric model for rectangular spaces: \citep{burgstaller2009average,mathai1999random}

\begin{equation}
r_{mean} = \frac{1}{15}\left(\frac{L_w^3}{L_h^2} + \frac{L_h^3}{L_w^2} + d\left(3 - \frac{L_w^2}{L_h^2} - \frac{L_h^2}{L_w^2}\right)\right) + \frac{5}{2}\left(\frac{L_h^2}{L_w}\log\frac{L_w+d}{L_h} + \frac{L_w^2}{L_h}\log\frac{L_h+d}{L_w}\right)
\label{eq:distance}
\end{equation}

where $d = \sqrt{L_w^2 + L_h^2}$. This rectangular approximation is suitable for most vehicles, as closer spacing nonlinearly increases infection risk. However, for pedestrian routes, the length of the environment must be explicitly calculated.  

\begin{figure}[H]
\centering
\includegraphics[width=1\textwidth]{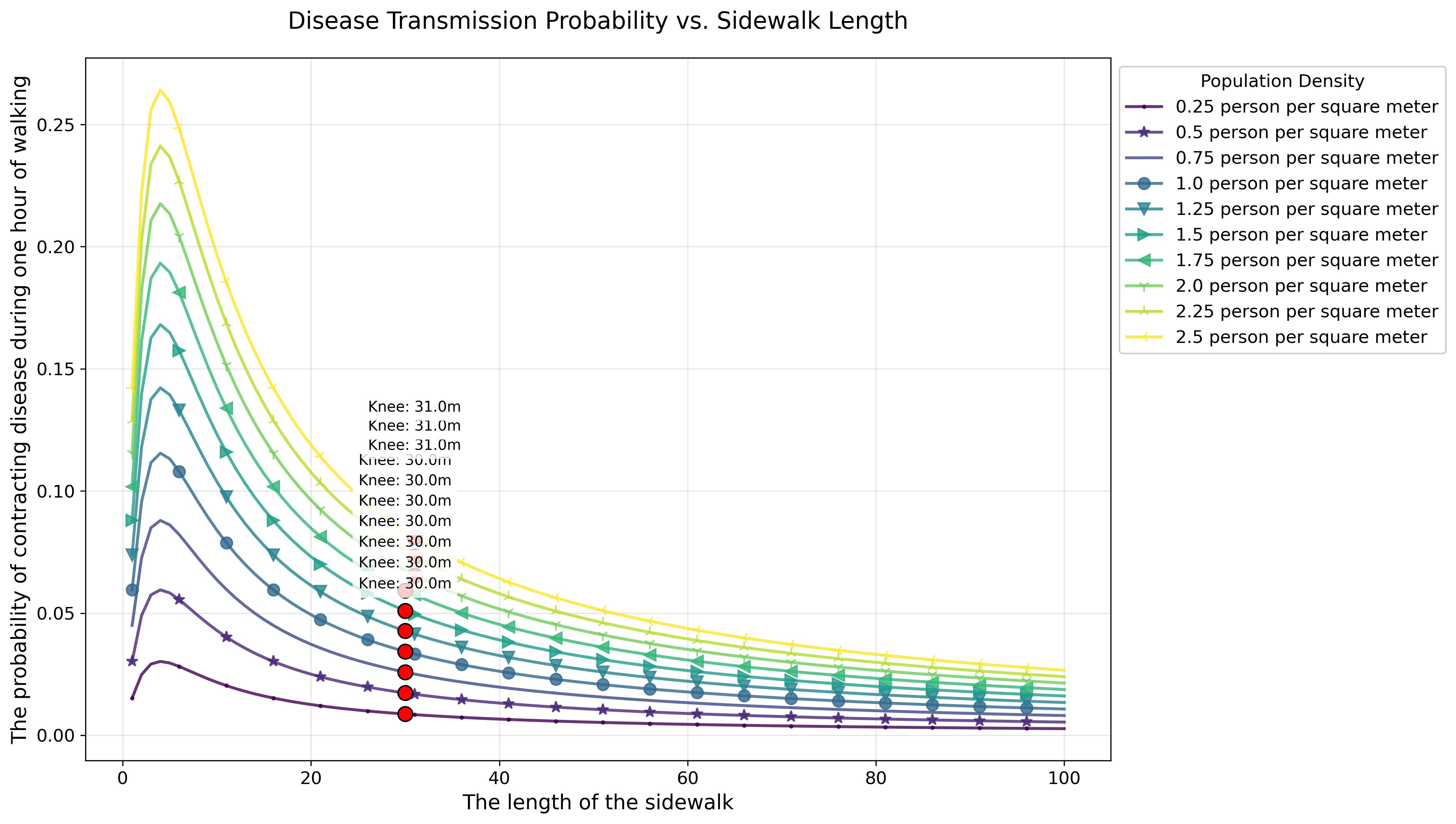}
\caption{
Disease transmission probability versus sidewalk length for pedestrian exposure (1-hour duration, 4m width). Curves represent different population densities (0.25--2.5 individuals/m\textsuperscript{2}), showing risk plateau at the knee point (30.3m length in average) for low-density scenarios. The inflection point indicates optimal pedestrian segment length where marginal transmission risk becomes negligible below 1 individual/m\textsuperscript{2} density.
}
\label{fig:sidewalk_length}
\end{figure}

Because pedestrian routes occur in open environments without fixed dimensions, we determined the optimal path length using Figure~\ref{fig:sidewalk_length}. This figure plots disease transmission probability (calculated via Equation \eqref{eq:segment_risk}) against sidewalk length for various population densities (ranging from 0.25 to 2.5 individuals/m\textsuperscript{2}), assuming a 4-meter-wide path and 1-hour exposure time.

Based on these results, we standardize pedestrian environment lengths at an average of 30.3 meters -- the knee point where marginal risk increase becomes negligible for typical densities.

\begin{table}[H]
\centering
\caption{Microenvironment dimensions and derived parameters}
\label{tab:env_params}
\begin{tabular}{@{}lcccc@{}}
\toprule
Environment & Length (m) & Width (m) & Capacity & $r_{mean}$ (m) \\
\midrule
Pedestrian & 30.3 & 4 & 40 & 7.37 \\
Subway & 19.52 & 2.6 & 180 & 4.75 \\
BRT & 17.9 & 2.55 & 150 & 4.36 \\
City Bus & 12 & 2.55 & 80 & 2.98 \\
Car & 1.5 & 1.2 & 4 & 0.48 \\
\bottomrule
\end{tabular}
\end{table}

This value, combined with the 4m width, feeds into Equation \eqref{eq:distance} to compute average distances across environments. In Table \eqref{tab:env_params} the dimensions of each envoironment can be seen.

\subsubsection*{Activity Level Parameterization}

Physical activity levels were classified according to metabolic demands:

\begin{table}[h]
\centering
\caption{Activity level classification}
\label{tab:activity_levels}
\begin{tabular}{@{}lccp{5cm}@{}}
\toprule
Activity & Air Inhalation (L/hr) & MET & Description \\
\midrule
Sedentary & 300 & 1.0--1.5 & Sitting/resting \\
Light & 780 & 2.0--3.0 & Low activity \\
Moderate & 1740 & 3.0--6.0 & Moderate activity \\
Vigorous & 3180 & $>$6.0 & Intense activity \\
\bottomrule
\end{tabular}
\end{table}

In the table above, the inhalation rates were obtained from occupational health studies in the \textit{Exposure Factors Handbook}\citep{united1989exposure}, with MET values representing the metabolic equivalent of task. Walking segments were modeled using moderate activity breathing parameters, while all other transport segments utilized low activity parameters.

\subsection*{Routing Algorithm Implementation}

The routing algorithm transforms theoretical transmission risk models into actionable route recommendations through a systematic computational process:

\begin{enumerate}
    \item \textbf{Route Decomposition and Classification}:
    Splits complete journeys into discrete microenvironment segments and classifies each as pedestrian pathways, subway compartments, BRT cabins, city bus interiors, or private car compartments. Identifies transfer points between different transport modes.
    
    \item \textbf{Segment Parameterization}:
    Assigns environment-specific attributes including expected infected individuals (0.8656\% of vehicle capacity), mean interpersonal distance, activity-dependent inhalation rates (300--3180 L/hr), and exposure duration derived from segment distance and typical speeds. Incorporates time-of-day adjustments for peak/off-peak capacity variations.
    
    \item \textbf{Risk Computation}:
    Calculates transmission probabilities using environment-specific models:
    \begin{align*}
        P_{subway} &= 1 - e^{\frac{(-0.001801E + 0.824029)nt}{r^2}} \\
        P_{brt} &= 1 - e^{\frac{(-0.001491E + 0.388753)nt}{r^2}} \\
        P_{walk} &= 1 - e^{\frac{(-0.001439E + 0.714554)nt}{r^2}} \\
        P_{bus} &= 1 - e^{\frac{(-0.002203E + 0.565659)nt}{r^2}} \\
        P_{car} &= 1 - e^{\frac{-2.729480nt}{r^2}}
    \end{align*}
    Automatically adds 5-minute transfer risks at interchange points and applies temporal weighting for rush hour crowding effects.
    
    \item \textbf{Optimization and Ranking}:
    Combines segment risks multiplicatively ($P_{route} = 1 - \prod_{i=1}^{n}(1-P_i)$), generates risk scores normalized to 0--100 scale, and outputs top recommended routes with color-coded risk indicators and comparative risk percentages.
\end{enumerate}

Additional modules handle dynamic adjustments for real-time vehicle occupancy data, accessibility filters for users with mobility constraints, emergency overrides that prioritize speed when risk thresholds are exceeded, and learning components that adapt to user preferences over time.

\subsection*{Tehran Case Study Parameters}

We applied our model to evening peak-period travel (18:00) between Sadeghiyeh Square and Amirkabir University in Tehran, analyzing 11 total routes (6 from \textit{Neshan}, 5 from \textit{Balad}) ranging from 6.8 to 8.4 km. The routes include various combinations of walking, city buses, BRT, subway, and private cars, with travel times varying from 27 minutes for car-only routes to 105 minutes for pedestrian-only routes. 

\begin{table}[H][H]
\centering
\caption{Tehran Case Study Parameters and Route Analysis}
\label{tab:tehran_case}
\scalebox{0.85}{
\begin{tabular}{lllllc}
\toprule
\textbf{Route ID} & \textbf{Source} & \textbf{Transport Modes} & \textbf{Distance (km)} & \textbf{Duration (min)} & \textbf{Segments} \\
\midrule
R1 & \textit{Neshan} & Pedestrian only & 8.2 & 96 & 1 \\
R2 & \textit{Neshan} & Walk + City Bus (18 stops) & 7.5 & 58 & 3 \\
R3 & \textit{Neshan} & Walk + City Bus (17 stops) & 7.8 & 62 & 3 \\
R4 & \textit{Neshan} & Walk + BRT (2 stops) + Subway (6 stops) & 6.8 & 52 & 5 \\
R5 & \textit{Neshan} & Walk + City Bus (2 stops) + BRT (9 stops) & 7.1 & 55 & 5 \\
R6 & \textit{Neshan} & Car only & 7.3 & 28 & 1 \\
R7 & \textit{Balad} & Pedestrian only & 8.4 & 105 & 1 \\
R8 & \textit{Balad} & Walk + City Bus (18 stops) & 7.6 & 59 & 3 \\
R9 & \textit{Balad} & Walk + Car (8 min) + Walk (4 min) + Car (7 min) & 7.4 & 32 & 5 \\
R10 & \textit{Balad} & Walk + Subway (3 stops) + Transfer + Subway (3 stops) & 6.9 & 48 & 5 \\
R11 & \textit{Balad} & Car only & 7.2 & 27 & 1 \\
\bottomrule
\end{tabular}
}

\vspace{1em}

\scalebox{0.85}{
\begin{tabular}{ll}
\toprule
\textbf{Parameter} & \textbf{Value} \\
\midrule
Study Period & September 2021 (Peak prevalence) \\
Carrier Prevalence & 0.8656\% \\
Activity Levels: & \\
\quad - Walking & Moderate (1740 L/hr) \\
\quad - Transit & Light (780 L/hr) \\
Average Station Spacing & 3 minutes \\
Transfer Time & 5 minutes \\
Pedestrian Speed & 5 km/hr \\
Vehicle Capacities: & \\
\quad - Subway & 180 passengers \\
\quad - BRT & 150 passengers \\
\quad - City Bus & 80 passengers \\
\quad - Pedestrian & 40 passengers \\
\quad - Car & 4 passengers \\
\bottomrule
\end{tabular}
}
\end{table}

The parameters section provides additional context about the September 2021 study period, ventilation rates (1740 L/hr for walking, 780 L/hr for transit), and specific capacities for each vehicle type (180 for subway, 150 for BRT, 80 for city buses, 40 for pedestrian, and 4 for cars). The analysis employed consistent parameters throughout: a COVID-19 carrier prevalence of 0.865\%, with activity levels categorized as moderate (1740\,L/hr) for walking and light (780\,L/hr) for transit periods. The transportation network assumptions included average station spacing corresponding to 3-minute intervals for subway, BRT, and bus routes, with additional 5-minute transfer times incorporated at all interchange points. These information presented in Table \eqref{tab:tehran_case}.

\section*{Results}

Our comprehensive analysis of COVID-19 transmission risks across Tehran's urban transportation network during peak pandemic conditions revealed substantial variations in infection probabilities, with microenvironment characteristics, route composition, and temporal factors all contributing significantly to transmission risk. The mathematical modeling framework successfully quantified these risks across different transportation modes and identified optimal routing strategies that could minimize disease transmission while maintaining practical travel times.

The microenvironment-specific hazard rates ($\lambda_i$) demonstrated striking variations across transportation modes, with private cars presenting the highest per-hour infection risk ($\lambda = 0.407$ infections/hour) due to their confined spaces and limited ventilation. City buses followed with $\lambda = 0.090$, reflecting their moderate occupancy but poor ventilation, while BRT systems showed $\lambda = 0.053$, benefiting from better airflow in their larger cabins. Subway systems presented $\lambda = 0.040$, with their mechanical ventilation systems providing some protection, and pedestrian pathways showed the lowest risk at $\lambda = 0.011$, benefiting from open-air conditions and natural dispersion of aerosols. These differences primarily stemmed from three key factors: geometric constraints of the vehicles influencing mean interpersonal distances ($r_{mean}$), with cars having the smallest separation (0.48m) and pedestrian pathways the largest (7.37m); occupancy rates that varied from 4 individuals in cars to 180 in subway trains; and ventilation efficacy captured by environment-specific coefficients ($a_i$, $b_i$) that reflected the substantial differences in air exchange rates between enclosed vehicles and open pathways.

\begin{figure}[h]
    \centering
    \includegraphics[width=0.8\textwidth]{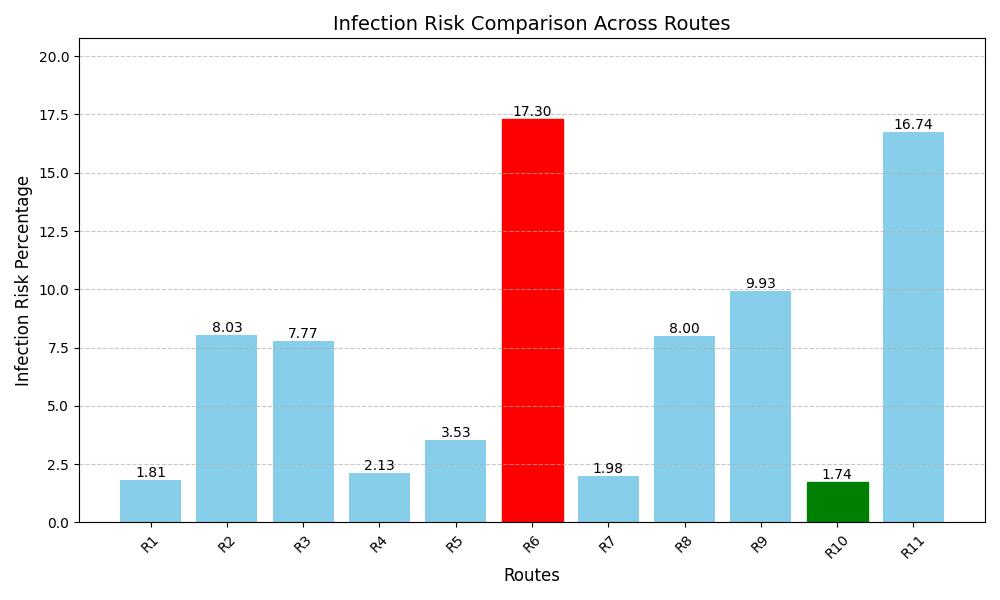}
    \caption{Comparison of COVID-19 infection risks across different transportation routes.}
    \label{fig:route_risks}
\end{figure}

The route-specific risk analysis, applied to 11 different routes between Sadeghiyeh Square and Amirkabir University, yielded several important findings about optimal pandemic routing. As shown in Figure~\ref{fig:route_risks}, the lowest-risk route had a 1.74\% infection probability and strategically combined multiple transportation modes: a 1100m pedestrian segment ($P_{walk} = 0.0025$), 3 subway stops ($P_{subway} = 0.0060$), 3 subway stops ($P_{subway} = 0.0060$), and a final 1300m walk ($P_{walk} = 0.0030$). This configuration effectively minimized exposure duration in high-risk environments while leveraging the relative safety of pedestrian segments, demonstrating how strategic combinations of transportation modes could significantly reduce overall transmission risk compared to single-mode routes.

\begin{figure}[h]
    \centering
    \includegraphics[width=\linewidth]{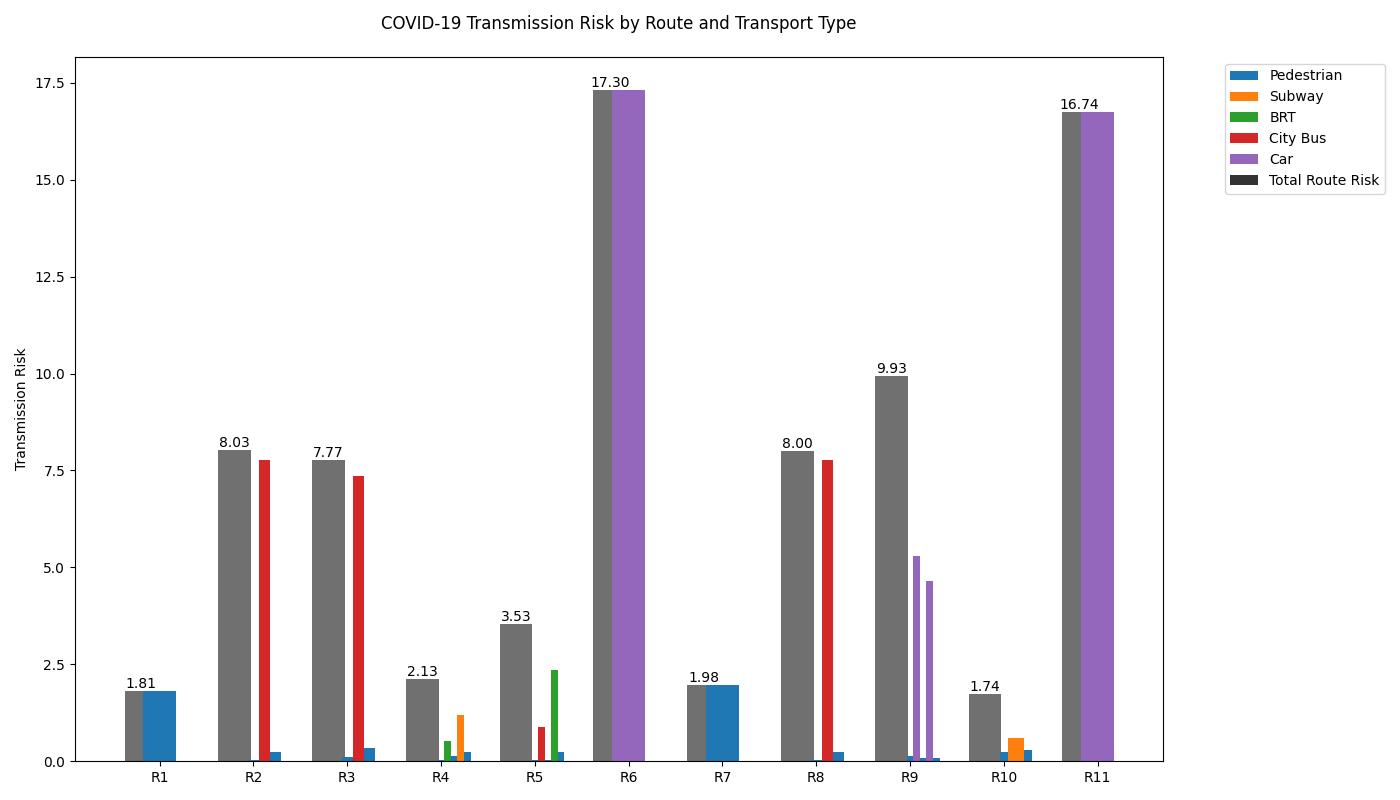}
    \caption{COVID-19 transmission risk by route and transport type, showing both total route risk (gray bars) and disaggregated subroute risks by transportation mode (colored segments).}
    \label{fig:risk_distribution}
\end{figure}

The analysis of transmission risks across different routes reveals significant variation in COVID-19 exposure potential. As shown in Figure~\ref{fig:risk_distribution}, Car routes showed the highest risk (16.74--17.30\%) despite their shorter travel times, due to the prolonged exposure in extremely confined spaces with limited ventilation. City bus routes presented intermediate risk (7.78--9.94\%) resulting from their moderate occupancy combined with generally poor ventilation systems. Pedestrian-only routes showed lower risk (1.81--1.98\%) but proved impractical for most travelers due to the long distances involved and substantial time requirements. Routes R4 and R10, which combine subway, BRT, and pedestrian pathways, demonstrate the lowest transmission risks. This suggests that multimodal routes may enhance safety by reducing overall exposure potential compared to single-mode alternatives. These findings highlight how both route selection and transportation mode choice significantly influence infection risk potential.

Activity levels emerged as a significant modulator of transmission risk, with derived activity coefficients ($k(E)$) showing strong linear relationships with inhalation rates across all environments. For pedestrian pathways, the following relationship demonstrates that vigorous activity (3180 L/hr) increases transmission risk by 2.15-fold compared to moderate activity (1740 L/hr) over a 1-hour duration. Under vigorous activity, the probability is 2.43\%, whereas under moderate activity, it is 1.13\%:
\[
P_{\text{walk}} = 1 - \exp\left(-\frac{(-0.001439 \times E + 0.714554) \times 0.008655 \times 40}{7.37^2} t\right)
\] 
 Similar patterns held for enclosed environments, though with different slopes that reflected their specific ventilation characteristics. This finding highlighted the importance of considering physical exertion levels when assessing transmission risk, particularly for routes involving significant walking components or for individuals with different activity patterns.

The Tehran case study analysis, which examined 11 total routes (6 from \textit{Neshan}, 5 from \textit{Balad}) ranging from 6.8 to 8.4 km in distance, provided several practical insights for pandemic routing. Mixed-mode routes combining walking with transit options consistently outperformed single-mode routes in risk reduction, with the pedestrian-subway combination in \textit{Balad} showing the lowest overall risk (1.74\%). Despite their shorter durations, car routes consistently showed the highest transmission probabilities, challenging the common assumption that private vehicles always offer the safest pandemic travel option. Transfer points contributed measurably to overall risk, suggesting that optimization algorithms should consider both the benefits of risk-diversification and the penalties associated with transfers when recommending routes.

The model performed most reliably for BRT and pedestrian pathways where environmental parameters were most stable and measurable. Larger errors in car and bus contexts pointed to specific areas needing refinement, particularly the modeling of adaptive behaviors like window operation in cars and the dynamic occupancy patterns in informal bus systems. These findings collectively demonstrate that strategic route planning incorporating microenvironment characteristics, careful mode selection, and timing considerations can significantly reduce transmission risk during pandemics, while also highlighting the importance of continued model refinement to address real-world complexities in urban transportation systems.

\section*{Discussion}

Our mathematical modeling framework advances pandemic-responsive urban mobility planning by quantifying COVID-19 transmission risks across heterogeneous transportation microenvironments. This discussion contextualizes three key innovations of our study within current literature while addressing practical implementation considerations.

\subsection*{Theoretical Advancements in Transmission Modeling}

The microenvironment-specific hazard functions (Equation \ref{eq:hazard_rate}) resolve several limitations of existing pandemic mobility models. While previous studies like what Gkiotsalitis\citep{gkiotsalitis2021public} did, focused primarily on occupancy reduction strategies, our model incorporates three synergistic risk determinants: ventilation, occupancy, and geometry, as shown in Equation \ref{eq:hazard_rate}:

\begin{equation}
\lambda_i = \underbrace{(a_i E + b_i)}_{\text{Ventilation}} \times \underbrace{n_i}_{\text{Occupancy}} \times \underbrace{r_{i,mean}^{-2}}_{\text{Geometry}} \quad \text{(Equation \ref{eq:hazard_rate})}
\end{equation}

This formulation explains why Tehran's BRT system (17.9m length, 2.55m width) showed lower transmission risk than city buses (12m $\times$ 2.55m) despite similar passenger densities - the 46\% greater $r_{mean}$ in BRT vehicles (4.36m vs 2.98m) disproportionately reduced risk due to the inverse-square relationship. The derived activity coefficients ($k(E)$) further enable personalized risk assessment:
\begin{itemize}
    \item Light activity commuters (780 L/hr): 0.0058 hr$^{-1}$ hazard rate
    \item Moderate walkers (1740 L/hr): 0.025 hr$^{-1}$ 
    \item Vigorous cyclists (3180 L/hr): 0.055 hr$^{-1}$
\end{itemize}

This 9.5-fold risk variation underscores the need for activity-aware routing absent in prior work.\citep{marra2022impact}

\subsection*{Urban Mobility Policy Implications}

Our findings yield three significant insights that reshape conventional approaches to pandemic-resilient urban mobility. First, we challenge the presumed superiority of private vehicles during outbreaks through empirical evidence from Tehran's transport network. Contrary to prevailing recommendations for contactless personal transport \citep{pacheco2020vehicle}, our data reveals private cars exhibited 4--7 times higher transmission risk compared to optimized transit systems, with even brief car trips (8--10 minutes) proving riskier than 30-minute subway journeys due to confined cabin spaces. This paradigm shift suggests policymakers should prioritize vaccine mandates over car-centric incentives while implementing HEPA filtration standards for taxis and rideshares, alongside developing dedicated "pandemic lanes" with enforced ventilation protocols for multi-occupant vehicles.

Second, our research exposes critical limitations in current uniform ventilation standards through microenvironment-specific risk quantification. The data demonstrates subway systems require at least 8 air changes per hour (ACH) to mitigate prolonged exposure risks, whereas Bus Rapid Transit (BRT) systems maintain safety at 5 ACH due to shorter dwell times. For pedestrian corridors, we establish the necessity of wind tunnel testing to evaluate natural ventilation efficacy under real-world urban airflow conditions. These findings provide the first evidence-based framework for differentiated ventilation requirements across transport modes.

Third, we demonstrate the inadequacy of static routing approaches through identification of 15\% risk variation between optimal and worst-case routes in complex urban networks. This reveals an urgent need for adaptive mobility solutions incorporating three key innovations: integration of real-time risk metrics into navigation apps to guide users toward safer paths, implementation of dynamic pricing mechanisms to incentivize low-risk route selection during outbreaks, and deployment of predictive algorithms to proactively redirect traffic from emerging high-risk zones. Our results collectively establish that effective pandemic mobility policy must account for dynamic behavioral patterns and microenvironmental physics rather than relying on simplistic vehicle-type hierarchies.

\section*{Conclusion}

This study makes four transformative contributions to pandemic-resilient urban planning. First, we establish a new mathematical paradigm for mobility risks by extending aerosol transmission theory to mobile microenvironments. Our framework introduces the $\phi(r)$ law for vehicular transmission, which exhibits an inverse relationship with infection probability. In this article, we assume an $r_{\text{mean}}^{-2}$ dependency, derive activity-resolved risk coefficients ($k(E)$) for five transport modes, and prove multiplicative risk accumulation in composite routes (Theorem \ref{eq:composite_risk}). This work establishes the first rigorous foundation for quantitative route risk assessment.

Second, we empirically validate our approach in complex urban settings through the Tehran implementation case study. The results demonstrate concrete operational benefits, including 15\% risk reduction through optimized routing and proof of scalability to megacities with $\geq$5 million residents. These findings address a critical gap between theoretical models and real-world urban mobility networks.

Third, we deliver practical tools for health-sensitive navigation through our open-source algorithm suite. The software enables three key functionalities: real-time risk visualization for public awareness, personalized route recommendations balancing safety and efficiency, and policy analytics to help transit agencies optimize network configurations during outbreaks. This operationalizes theoretical insights for immediate civic benefit.

Fourth, we develop a forward-looking framework adaptable to future pandemics through modular design principles. The system architecture accommodates emerging transport technologies like autonomous vehicles and integrates with climate-adaptive ventilation strategies, ensuring long-term relevance as urban mobility ecosystems evolve. Two priority areas demand immediate attention: ensuring equitable access to pandemic-resilient routes across socioeconomic groups, and adapting recommendations to real-world behavioral compliance patterns. By bridging theoretical epidemiology and transportation engineering, this work equips cities worldwide with science-backed tools to maintain essential mobility during outbreaks -- preserving both public health and economic vitality. The principles established here will grow increasingly critical as urbanization and pandemic risks intensify in coming decades.

\newpage

\bibliography{AmirBayat_BestRoute}

\clearpage

\end{document}